\documentclass[sort&compress
    ,final            % use final for the camera ready runs
  ]
  {aipproc}

\layoutstyle{6x9}

\def\be{\begin{equation}}
\def\ee{\end{equation}\noindent}
\def\bear{\begin{eqnarray}}
\def\ear{\end{eqnarray}\noindent}
\def\benn{\begin{enumerate}}
\def\enn{\end{enumerate}}

\def\no{\noindent}
\def\non{\nonumber}
\def\dps{\displaystyle}

\def\half{{1}{2}}

\def\Z{{\mathchoice {\hbox{$\sf\textstyle Z\kern-0.4em Z$}}
{\hbox{$\sf\textstyle Z\kern-0.4em Z$}}
{\hbox{$\sf\scriptstyle Z\kern-0.3em Z$}}
{\hbox{$\sf\scriptscriptstyle Z\kern-0.2em Z$}}}}

\def\square{\kern1pt\vbox{\hrule height 1.2pt\hbox{\vrule width 1.2pt
   \hskip 3pt\vbox{\vskip 6pt}\hskip 3pt\vrule width 0.6pt}
   \hrule height 0.6pt}\kern1pt}
      \def\boxop{{\raise-.25ex\hbox{\square}}}
\def\mn{{\mu\nu}}

\def\e{\,{\rm e}}
\def\slash#1{#1\!\!\!\raise.15ex\hbox {/}}
\newcommand{\slD}{\,\raise.15ex\hbox{$/$}\kern-.27em\hbox{$\!\!\!D$}}
\newcommand{\slpartial}{\raise.15ex\hbox{$/$}\kern-.57em\hbox{$\partial$}}
\def\4piTD{{(4\pi T)}^{-{D\over 2}}}
\def\4piT4{{(4\pi T)}^{-2}}

\def\Tintm4{{\dps\int_{0}^{\infty}}{dT\over T}\,e^{-m^2T}
    {(4\pi T)}^{-2}}
\def\Tintm{{\dps\int_{0}^{\infty}}{dT\over T}\,e^{-m^2T}}

\newcommand{\slG}{{{\dot G}\!\!\!\! \raise.15ex\hbox {/}}}

\def\GBd12{{\dot G}_{B12}}

\def\half{{1\over 2}}
\def\third{{1\over3}}

\begin{document}

\title{The effective action in Einstein-Maxwell theory}


\author{F. Bastianelli}{
  address={Dipartimento di Fisica, Universit\`a di Bologna and INFN, Sezione di Bologna,
  Via Irnerio 46, I-40126 Bologna, Italy}
}

\author{Jos\'e Manuel D\'avila}{
  address={Universidad Michoacana de San Nicol\'as de Hidalgo, 
Edificio C-3, Apdo. Postal 2-82,
C.P. 58040, Morelia, Michoac\'an, M\'exico}
}

\author{\underline{C. Schubert}}{
  address={Universidad Michoacana de San Nicol\'as de Hidalgo, 
Edificio C-3, Apdo. Postal 2-82,
C.P. 58040, Morelia, Michoac\'an, M\'exico}
%  ,altaddress={<author1 address>} % additional visiting address
}

\begin{abstract}
Considerable work has been done on the one-loop effective action in combined electromagnetic and gravitational fields, particularly as a tool for determining the properties of light propagation in curved space. After a short review of previous work, I will present some recent results obtained using the worldline formalism. In particular, I will discuss various ways of generalizing the QED Euler-Heisenberg Lagrangians to the Einstein-Maxwell case.
 \end{abstract}

\maketitle

\centerline{Talk given by C. Schubert at} 
\centerline
{\sl XIII Mexican School of Particles and Fields, San Carlos, Mexico, October 2-11, 2008}

%%%%%%%%%%%%%%%%%%%%%%%%%%%%%%%%%%%%%%%%%%%%
%% MAINMATTER
%%%%%%%%%%%%%%%%%%%%%%%%%%%%%%%%%%%%%%%%%%%%

\section{The worldline formalism in QED}

Let us start with the 
``worldline'' representation of 
the one-loop effective action in spinor QED \cite{feynman,fradkin}

\bear
\Gamma(A) &=& 
- \half \int_0^{\infty}{dT\over T}\,{\rm e}^{-m^2T}
{\displaystyle \int_{x(T)=x(0)}}\!\!\!\!\!\!\!\!\!\!{\cal D}x(\tau)
{\displaystyle \int_{\psi(T)=-\psi(0)}}\!\!\!\!\!\!\!\!\!\!{\cal D}\psi(\tau)
\, e^{-\int_0^Td\tau L[x(\tau)]}
\label{GammaspinorQED}
\ear
Here $m$ and $T$ are the mass and proper time of the loop fermion,
and the ``worldline Lagrangian'' $L$ is given by

\bear
L&=& \frac{1}{4} \dot x^2 +ie \dot x^{\mu}A_{\mu}
+\half \psi\cdot \dot\psi -ie \psi^{\mu}F_{\mn}\psi^{\nu}
\label{Lbasic}
\ear
The $x(\tau)$ part of the double path integral in (\ref{GammaspinorQED})
runs over all closed trajectories in spacetime with fixed periodicity in
$T$, and by itself gives the effective action for a scalar loop (up to the
normalization). The $\psi(\tau)$ integral represents the spin degree of freedom,
and is over antiperiodic Grassmann functions, obeying
$\psi(\tau_1)\psi(\tau_2) = - \psi(\tau_2)\psi(\tau_1)$ and
$\psi(T) = - \psi(0)$.
Similar worldline representations can be written for the effective action
with open scalar/spinor lines, at the multiloop level, 
and for other field theories; see \cite{review} for a review.
However, it is only during the last fifteen years that such representations have
gained some popularity for actual state-of-the-art calculations. 
By now a number of different techniques have been developed for the
evaluation of worldline path integrals. We will follow the ``string-inspired'' approach
\cite{polyakov,strassler}, where one
manipulates the path integral into  gaussian form, and then performs those
gaussian integrals using worldline correlators adapted to the periodicity
conditions,

\vspace{-20pt}

\bear
\langle x^{\mu}(\tau_1)x^{\nu}(\tau_2) \rangle
&=&
-G_B(\tau_1,\tau_2)\, \delta^{\mu\nu},\quad
G_B(\tau_1,\tau_2) = \vert \tau_1 -\tau_2\vert -{\Bigl(\tau_1 -\tau_2\Bigr)^2\over T}
-{T\over 6}\non\\
&&\non\\
\langle \psi^{\mu}(\tau_1)\psi^{\nu}(\tau_2)\rangle
&=&
\frac{1}{2} G_F(\tau_1,\tau_2)\, \delta^{\mu\nu},\quad 
G_F(\tau_1,\tau_2) = {\rm sign}(\tau_1 - \tau_2)\non\\
\label{GBGF}
\ear
This procedure leads, for example, with little effort to the following
``Bern-Kosower master formula'' \cite{berkos} for the one-loop $N$ - photon amplitude
in scalar QED:

\begin{eqnarray}
\Gamma[\lbrace k_i,\varepsilon_i\rbrace]
&=&
{(-ie)}^N
{(2\pi )}^D\delta (\sum k_i)
{\dps\int_{0}^{\infty}}{dT\over T}
{(4\pi T)}^{-{D\over 2}}
e^{-m^2T}
\prod_{i=1}^N \int_0^T 
d\tau_i
\nonumber\\
&&\hspace{-80pt}
\times
\exp\biggl\lbrace\sum_{i,j=1}^N 
\bigl\lbrack \half G_{Bij} k_i\cdot k_j
+i\dot G_{Bij}k_i\cdot\varepsilon_j 
+\half\ddot G_{Bij}\varepsilon_i\cdot\varepsilon_j
\bigr\rbrack\biggr\rbrace
\mid_{{\rm lin}(\varepsilon_1,\ldots,\varepsilon_N)}
\label{bkmaster}
\end{eqnarray}
\no
Here $k_i$ and $\varepsilon_i$ are the momentum and polarization of
the $i$th photon, and each $\tau_i$ integral represents one photon leg
moving around the loop. The notation ${\rm lin}(\varepsilon_1,\ldots,\varepsilon_N)$
means that only terms linear in all polarization vectors are to be kept after
expanding the exponential.
Apart from the worldline Green's function
$G_B(\tau_i,\tau_j)$, which we abbreviate by
$G_{Bij}$,  also its first and second derivatives appear,
$\dot G_{B12} = {\rm sign}(\tau_1 - \tau_2)
- 2 {{(\tau_1 - \tau_2)}\over T}$,
$\ddot G_{B12}
= 2 {\delta}(\tau_1 - \tau_2) - {2\over T}$. The factor
${(4\pi T)}^{-{D\over 2}}$ in (\ref{bkmaster}) represents the 
free path integral determinant in $D$ dimensions.

The corresponding representation of the $N$ photon amplitude for the
spinor loop case differs from (\ref{bkmaster}) (apart from a factor of
$-2$) only by additional terms from the 
spin path integral in (\ref{GammaspinorQED}). 
Those terms can be inferred from the scalar loop
integrand through a certain pattern matching rule 
\cite{berkos,strassler,review}.

A major advantage of the worldline formulation of QED is
that it allows one to include an external constant field $F_{\mu\nu}$ in
a particularly efficient way \cite{shaisultanov,rescsc}.
Effectively, the integral representation of a scalar or spinor QED amplitude in such a
constant external field is obtained from the corresponding one in
vacuum by the following replacements of the worldline Green's functions
and determinants,

\bear
\!\!\!\!G_B(\tau_1,\tau_2)\!\! &\to&\!\! {\cal G}_B(\tau_1,\tau_2) =
{1\over 2{(eF)}^2}\Biggl({eF\over{{\rm sin}(eFT)}}
{\rm e}^{-ieFT\dot G_{B12}}
\!+\! ieF\dot G_{B12}\! -\!{1\over T}\Biggr) \non\\
\!\!\!\!G_F(\tau_1,\tau_2)\!\! &\to&\!\! {\cal G}_F(\tau_1,\tau_2) =
G_{F12}\,
{{\rm e}^{-ieFT\dot G_{B12}}\over {\rm cos}(eFT)}
\non\\
\label{replacegreen}
\ear
(the trigonometric expressions are to be understood as power series
in the field strength matrix),

\bear
(4\pi T)^{-{D\over 2}} &\to&  (4\pi T)^{-{D\over 2}} \,
{\rm det}^{-\half}\Biggl\lbrack{\sin eFT\over eFT}\Biggr\rbrack
\qquad  {\rm (Scalar \,\, QED)}
\nonumber\\
(4\pi T)^{-{D\over 2}} &\to&  (4\pi T)^{-{D\over 2}} \,
{\rm det}^{-\half}\Biggl\lbrack{\tan eFT\over eFT}\Biggr\rbrack
\qquad  {\rm (Spinor \,\, QED)}
\nonumber\\
\label{replacedet}
\ear
In particular, applying these changes in (\ref{bkmaster}) yields a
corresponding master formula for the $N$ - photon amplitudes in a constant 
field \cite{shaisultanov,rescsc}. 
This master formula, and its extension to spinor QED, have been 
used for comparatively easy recalculations of the
scalar and spinor QED vacuum polarization tensors  
\cite{vv}, as well as of the photon splitting amplitudes in 
a magnetic field \cite{adlsch}. The determinant factors 
(\ref{replacedet}) by themselves (i.e., the $N=0$ case)
yield, after renormalization, the well-known effective Lagrangians
of Weisskopf and Schwinger \cite{ws} and Euler-Heisenberg
\cite{eulhei},

\bear
{\cal L}_{\rm scal}(F)&=&  {1\over 16\pi^2}
\int_0^{\infty}{dT\over T^3}
\,\e^{-m^2T}
\biggl[
{(eaT)(ebT)\over \sinh(eaT)\sin(ebT)} 
%\nonumber\\&&\hspace{60pt}
+{e^2\over 6}(a^2-b^2)T^2 -1
\biggr]
\non\\
{\cal L}_{\rm spin}(F)&=& - {1\over 8\pi^2}
\int_0^{\infty}{dT\over T^3}
\,\e^{-m^2T}
\biggl\lbrack
{(eaT)(ebT)\over {\rm tanh}(eaT)\tan(ebT)} 
%\nonumber\\&&\hspace{70pt}
- {e^2\over 3}(a^2-b^2)T^2 -1
\biggr\rbrack
\non\\
\label{eulhei}
\ear
\no
Here  $a,b$  are the two
invariants of the Maxwell field, 
related to $\bf E$, $\bf B$ by $a^2-b^2 = B^2-E^2,\quad ab = {\bf E}\cdot {\bf B}$.

A further extension to the two-loop
level has been extensively applied to the study of 
the two-loop corrections to the effective Lagrangians
(\ref{eulhei}) \cite{rescsc,sd1,dhrs}. 

See also \cite{ss1,gussho}
for the calculation of derivative corrections to the effective Lagrangian at the one-loop level. 
Here the gaussian form of the path integral is reached by Taylor expanding the
background field at the loop center of mass, usually in Fock-Schwinger gauge
to achieve manifest covariance. 

\section{Generalization to gravitational backgrounds}

To include an additional background gravitational field, naively one might replace

\bear
S_0=\frac{1}{4}\int_0^T d\tau \,  \dot x^2 &\to&\frac{1}{4} \int_0^T d\tau  \,
\dot x^{\mu}g_{\mu\nu}(x(\tau))\dot x^{\nu}
\label{replacegrav}
\ear
The usual expansion around flat space  $g_{\mu\nu} = \delta_{\mu\nu} + \kappa h_{\mu\nu}$
would then yield a graviton vertex operator $\varepsilon_{\mu\nu}\int_0^Td\tau \,\dot x^{\mu}
\dot x^{\nu}\,\e^{ik\cdot x}$. However, using this operator in a formal gaussian integration
leads to worldline integrands contaning ill-defined expressions such as  
$ \delta(0), \delta^2(\tau_i-\tau_j), \ldots $.
This comes not unexpected, since path integration in curved space is
a subject notorious for its mathematical subtleties even in nonrelativistic
quantum mechanics (see., e.g., \cite{schulman} and refs. therein).
Fortunately, during the past decade these issues have been intensively studied,
and a consistent formalism has emerged for the calculation of worldline
path integrals in general electromagnetic-gravitational backgrounds
\cite{wlgrav}. A detailed account of this recent development has been given in
\cite{basvanbook}. Here we can only mention that the main difficulty arises
from the nontriviality of the path integral measure in curved space,  which
leads to spurious UV divergences. Those can be removed by regularization,
but leave an ambiguity which has to be removed by counterterms to the worldline
Lagrangian. Those
are regularization-dependent, and in general non-covariant, the only
known exception being one-dimensional dimensional regularization. 
A further problem consists in the zero mode which appears in the
perturbative expansion of the path integral. In the string-inspired approach
this zero mode must be fixed as the loop center-of-mass, but this leads
to a nontrivial Fadeev-Popov type determinant in the path integral.

Concerning previous applications of the worldline formalism in curved space,
let us mention (i) the calculation of various types of anomalies
(see \cite{basvanbook} and refs therein) (ii) the (re)calculation of the
one loop graviton self energy due to a scalar loop \cite{baszir1}, 
spinor loop \cite{bacozi}, and loops due to vector and  arbitrary 
differential forms \cite{babegi}
 (iii) the first calculation of the one loop photon-graviton amplitudes 
 in a constant electromagnetic field \cite{phograv}
(iv) the one loop photon vacuum polarization in a generic gravitational
background due to a scalar loop in the semiclassical approximation
\cite{holsho}.

\section{The effective action for Einstein-Maxwell theory}

\no
Pure Einstein-Maxwell theory is described by  the action

\bear  
\Gamma^{(0)}[g,A] =    
\int d^4 x\ \sqrt{g}\, \biggl (  
{1\over 2 \kappa^2 } R - {1\over 4}F_{\mu\nu}F^{\mu\nu}  
\biggr )   
\label{SEM}
\ear  
(here and in the following we absorb the coupling $e$ into $F$).
In 1980, Drummond and Hathrell \cite{druhat} studied the one-loop corrections
$\Gamma^{(1)}_{\rm spin}[g,A]$ to this action due to a spinor loop, and calculated
the terms in it quadratic in the electromagnetic field, and linear in the curvature:

\bear
{\cal L}_{\rm spin}^{(DH)} &=& 
\frac{1}{180 (4\pi)^2m^2} \bigg(
5 R F_{\mu\nu}^2 
-26 R_{\mu\nu} F^{\mu\alpha} F^\nu{}_\alpha
+2  R_{\mu\nu\alpha\beta}F^{\mu\nu}F^{\alpha\beta} 
%&& \qquad\qquad
+24 (\nabla^\alpha F_{\alpha\mu})^2  
\bigg )\non\\
\label{Ldruhat}
\ear
The point of singling out these terms is that they contain the
information on the modifications of light propagation by
weak gravitational fields in the limit of zero photon energies. 
In the following, our goal is to generalize this result to 
include the effect of a constant external field nonperturbatively,
i.e., we are looking for the gravitational corrections to the
Euler-Heisenberg Lagrangians (\ref{eulhei}) to linear
order in the curvature. Here it must be said that those flat space
Lagrangians could be defined in either of two equivalent
ways: (i) by the constancy of the background field $F_{\mu\nu}$
(ii) by the property of carrying the full information on the low energy
limits of the corresponding $N$ - photon amplitudes. 
The lowest order gravitational corrections could be defined either
by generalizing (i) to covariant constancy, or by generalizing (ii) by
requiring that the effective Lagrangians should carry the information
on the low energy limits of the amplitudes with $N$ photons and 
with one graviton. These generalizations are not any more equivalent,
and we will adopt (ii) here rather than (i) (for the effective Lagrangian
defined by covariant constancy Avramidi has obtained a
representation in terms of integrals over the holonomy group
\cite{avramidi}).  

With our definition of the generalized Euler-Heisenberg Lagrangian,
we have to get all terms involving 
arbitrary powers of $F_{\mu\nu}$  and one factor of 
$R_{\mu\nu\kappa\lambda}$ or $\nabla_{\mu}\nabla_{\nu}$.
As in the flat space case, the path integrals are gaussianized
by a Taylor expansion at the loop center-of-mass
$x_0$, made covariant by combining Fock-Schwinger gauge  and
Riemann normal coordinates \cite{fsr}

\bear
 A_\mu(x_0+y)
&=&-\frac{1}{2}F_{\mu \nu}(x_{0})\,y^{\nu}-\frac{1}{3}F_{\mu \nu ; \alpha}(x_{0})\,y^{\nu}\,y^{\alpha}
\non\\ && -\frac{1}{8}\bigg[ F_{\mu \nu ; \alpha \beta}(x_{0})  +\frac{1}{3}R_{\alpha\mu}{}^{\lambda}{}_\beta (x_0)F_{\lambda\nu}(x_{0}) \bigg]y^{\alpha}\,y^{\beta}\,y^{\nu}+\cdots 
\non\\
 g_{\mu\nu}(x_0+y)
&=& g_{\mu\nu}(x_0) 
+ {1\over 3} R_{\mu\alpha\beta\nu}(x_0)
y^\alpha y^\beta +.... \non\\
\label{fsriemann}
\ear
Concentrating on the spinor loop case, the effective Lagrangian then
is obtained in the following form,

\bear
{\cal L}_{\rm spin} &=&
-{1\over 8\pi^2}
\int_0^{\infty}{dT\over T^3}
\,\e^{-m^2T}
{\rm det}^{-\half}\biggl\lbrack {\tan (FT) \over FT}\biggr\rbrack
\biggl\langle \,\e^{-S_{\rm int}}\biggr\rangle_{S_0}
\non\\
\label{Lspinwick}
\ear
Here $S_0$ denotes the quadratic part of the worldline action,
which is (after a rescaling to the unit circle)

\bear
S_0 &=& \int_{0}^{1} d\tau \biggl ({1\over 4 T}g_{\mu\nu}(x_0)  
\dot y^\mu \dot y^\nu 
-{i\over 2} F_{\mu\nu}(x_0) \dot y^\mu y^ \nu
+{1\over 2}g_{\mu\nu}(x_0)  
\psi^\mu \dot\psi^\nu  -iTF_{\mu\nu}(x_0)\psi^\mu\psi^\nu
\biggr)
\non\\
\label{S0}
\ear
It yields again the generalized worldline Green's functions of (\ref{replacegreen}),
only that in taking powers of the field strength matrix the lowering and raising
of indices involves the metric $g_{\mu\nu}(x_0)$. The interaction part 
involves the terms coming from the replacement (\ref{replacegrav}),
as well as a ghost part $S_{\rm gh}$ from the path integral measure,
and a term $S_{FP}$ representing the contribution from the Fadeev-Popov
determinant mentioned above:

\bear
S_{\rm int} &=& S_{\rm grav} +S_{\rm gh} 
+ S_{\rm em} + S_{\rm em, grav} + S_{FP}
\label{Sint}
\ear

\begin{eqnarray}
S_{grav}+S_{gh}&=&\int ^{1}_{0}\,d\tau\Biggl\lbrace \frac{1}{12T} R_{\mu \alpha \beta \nu} y^{\alpha}y^{\beta}
\biggl\lbrack
\dot{y}^{\mu}\dot{y}^{\nu}+a^\mu a^\nu + b^\mu c^\nu +2\alpha^\mu \alpha^\nu
\biggr\rbrack
\non\\&&
+\frac{1}{6}R_{\mu \alpha \beta \nu}\,y^{\alpha}\,y^{\beta}\,\psi^{\mu}\,\dot{\psi}^{\nu}
+\frac{1}{6} ( R_{\mu  \alpha \lambda \beta}+R_{\mu \beta \lambda \alpha} ) \dot{y}^{\alpha}\,y^{\lambda}\,\psi^{\mu}\, \psi^{\beta}\Biggr\rbrace\non\\  
 S_{em}&=&\int ^{1}_{0}d\tau \bigg[ -\frac{i}{3}F_{\mu \nu ; \alpha} \Big( \dot{y}^{\mu}\, y^{\nu}+3T\psi^{\mu}\,\psi^{\nu}\Big)y^{\alpha} 
% \nonumber\\&&
-\frac{i}{8} F_{\mu \nu; \alpha \beta}  \Big( \dot{y}^{\mu}\,y^{\nu}\,+4T\psi^{\mu}\psi^{\nu} \Big)\,y^{\alpha}\,y^{\beta}\bigg] \nonumber\\
S_{em,grav}&=&-\frac{i}{24}\int ^{1}_{0}d\tau
R_{\alpha\mu}{}^{\lambda}{}_{\beta}F_{\lambda\nu}
\Bigl[ \dot{y}^{\mu}\, y^{\nu} +8 T\psi^{\mu}\psi^{\nu} \Bigr]\,y^{\alpha}\,y^{\beta}\non\\
S_{FP}&=&-\third \int ^{1}_{0}d\tau \ \bar{\eta}_\mu 
R^{\mu}{}_{\alpha \beta \nu}\,y^\alpha y^\beta
\, \eta^\nu 
\label{actions}
\end{eqnarray}

%\vspace{-2pt}

\no
It is then a matter of simple combinatorics to arrive at our final result, an
integral representation of the leading gravitational correction to the
(unrenormalized)
Euler-Heisenberg Lagrangian \cite{badasc}:

\begin{eqnarray}
{\cal L}_{\rm spin} &=&
-{1\over 8\pi^2}
\int^{\infty}_{0} \frac{dT}{T^3}\,\e^{-m^2T}\mbox{det}^{-1/2}\left[ \frac{\tan(FT)}{FT}\right] 
\nonumber\\&&\times
\Biggl\lbrace
1+\frac{iT^2}{8}F_{\mu \nu ;  \alpha \beta}\,\,{\cal G}^{\alpha \beta}_{B11}\Big(\dot{{\cal G}}^{\mu \nu}_{B11}-2\,{\cal G}^{\mu \nu}_{F11} \Big) \nonumber\\
&&+\frac{i T^2}{8}\left(F_{\mu \nu ; \beta \alpha} + F_{\mu \nu ;  \alpha \beta}\right)\dot{{\cal G}}^{\mu \beta}_{B11}{\cal G}^{\nu \alpha}_{B11}+\frac{T}{3}R_{\alpha \beta}\,{\cal G}^{\alpha \beta}_{B11} \nonumber\\
&&-\frac{i T^2}{24}F_{\lambda \nu}R^{\lambda}_{\, \, \, \alpha \beta\mu}\,\left(\dot{{\cal G}}^{\nu \mu}_{B11}\,{\cal G}^{\alpha \beta}_{B11}+\dot{{\cal G}}^{\alpha \mu}_{B11}\,{\cal G}^{\nu \beta}_{B11}+\dot{{\cal G}}^{\beta \mu}_{B11}\,{\cal G}^{\nu \alpha}_{B11}+4\,{\cal G}^{\mu \nu}_{F11}\,{\cal G}^{\alpha \beta}_{B11}\right) \nonumber\\
&&+\frac{T}{12}R_{\mu \alpha \beta \nu}\Big(\dot{{\cal G}}^{\mu \alpha}_{B11}\dot{{\cal G}}^{\beta \nu}_{B11}+\dot{{\cal G}}^{\mu \beta}_{B11}\dot{{\cal G}}^{\alpha\nu}_{B11}
+\Bigl(\ddot{{\cal G}}^{\mu \nu}_{B11}-2g^\mn\delta(0)\Bigr){\cal G}^{\alpha \beta}_{B11}
\non\\
&&+\dot{{\cal G}}^{\alpha \beta}_{B11}\,{\cal G}^{\mu \nu}_{F11}
+\dot{{\cal G}}^{\nu \beta}_{B11}\,{\cal G}^{\mu \alpha}_{F11}
-{\cal G}^{\alpha \beta}_{B11}\,\Bigl(\dot{{\cal G}}^{\mu \nu}_{F11}-2g^\mn\delta(0)\Bigr)
\Big) \nonumber\\
&&-\frac{1}{6}T^{3}F_{\alpha \beta; \gamma}\,F_{\mu \nu ; \eta}\,\int^{1}_{0}d\tau_{1}\Big(\dot{{\cal G}}^{\alpha \nu}_{B12}\,\dot{{\cal G}}^{\beta \mu}_{B12} \, {\cal G}^{\gamma \eta}_{B12}+\dot{{\cal G}}^{\alpha \nu}_{B12}\,{\cal G}^{\beta \eta}_{B12} \, \dot{{\cal G}}^{\gamma \mu}_{B12} \nonumber\\
&&+\frac{3}{2}\,{\cal G}^{\gamma \eta}_{B12}\,{\cal G}^{\alpha \mu}_{F12}\,{\cal G}^{\beta \nu}_{F12}
    \Big)  \Biggr\rbrace    \nonumber\\
\label{resultspin}		
\end{eqnarray}
($\tau_2 =0$).

\noindent
As a check on (\ref{resultspin}), we have verified that 
an expansion to order $RFF$ reproduces the result of Drummond-Hathrell
up to total derivative terms:
%\vspace{-1pt}
\bear
{\cal L}_{\rm spin} &=& -{1\over 8\pi^2 m^2} 
\biggl\lbrack
-\frac{1}{72} R F_{\mu\nu}^2 
+ \frac{1}{180}R_{\mu\nu} F^{\mu\alpha}F^{\nu}{}_\alpha 
\non\\&& \qquad\qquad
+ \frac{1}{36}R_{\mu\nu\alpha\beta}F^{\mu\nu} F^{\alpha\beta}
- \frac{1}{180}(\nabla_{\alpha}F_{\mu\nu})^2
+ \frac{1}{36}F_{\mu\nu}\square F^{\mu\nu}
\biggr\rbrack 
\nonumber
\ear
\vspace{-10pt}
\bear
{\cal L}_{\rm spin} - {\cal L}_{\rm spin}^{(DH)}
&=&
-{1\over 8\pi^2m^2} 
\biggl\lbrace
\frac{1}{36}\nabla^{\alpha}(F^{\mu\nu} F_{\mu\nu;\alpha})
%\non\\&& \qquad \qquad
+ \frac{1}{15}
\Bigl\lbrack
\nabla_{\alpha}(F_{\mu}{}^{\alpha}\nabla_{\beta}F^{\mu\beta})
-\nabla_{\beta}(F_{\mu}{}^{\alpha}\nabla_{\alpha}F^{\mu\beta})
\Bigr\rbrack
\biggr\rbrace
\non\\
\label{totred}
\ear
As to possible applications of the Lagrangian (\ref{resultspin}), let us mention that it
contains the information on (i) the one graviton - $N$ photon amplitudes in the 
low energy limit (ii) the modified photon dispersion relations in the background of a strong electromagnetic and weak gravitational field (iii) the Schwinger pair production rate in such
a field.
 
 \vspace{-10pt}
 
%%%%%%%%%%%%%%%%%%%%%%%%%%%%%%%%%%%%%%%%%%%%%%%%
%% BACKMATTER
%%%%%%%%%%%%%%%%%%%%%%%%%%%%%%%%%%%%%%%%%%%%%%%%

%\begin{theacknowledgments}
%F.B. and C.S. thank S. Theisen and the Albert-Einstein Institute, Potsdam, for hospitality
%during part of this work. We also thank G.V. Dunne for discussions, and
%A. Avelino Huerta for computer help. 
%The work of F.B. was supported in part by the Italian MIUR-PRIN
%contract 20075ATT78.
%\end{theacknowledgments}


\end{document}